# Mechanical manipulation and exfoliation of boron nitride flakes by micro-plowing with an AFM tip

Joshua O. Island[1], Gary A. Steele[1], Herre S.J. can der Zant[1], Andres Castellanos-Gomez[1,2*]

[1] Kavli Institute of Nanoscience, Delft University of Technology, Lorentzweg 1, 2628 CJ Delft, The Netherlands.

[2] Present address: Instituto Madrileño de Estudios Avanzados en Nanociencia (IMDEA-Nanociencia), 28049 Madrid, Spain.

[*] Electronic mail: a.castellanosgomez@tudelft.nl or andres.castellanos@imdea.org .

## ABSTRACT

We demonstrate a simple method to manipulate and exfoliate thick hexagonal boron nitride (h-BN) flakes using an atomic force microscope (AFM) cantilever and tip mounted to a micromanipulator stage. Thick flakes of tens to hundreds of nanometers can be thinned down to large area, few-layer and monolayer flakes. We characterize the resulting thinned-down flakes using AFM and Raman spectroscopy. This technique is extendable to other two dimensional, van der Waals materials.

## MAIN TEXT

### Introduction

Mechanical exfoliation has provided a powerful, yet simple method to fabricate 2D materials [1-6]. This method, which can be carried out in most physics or material science laboratories, is fast, inexpensive and provides high quality samples. The main limitation of this technique is the lack of control during the mechanical exfoliation process as atomically thin flakes are transferred at random positions on the substrate. Moreover thick flakes are also transferred making necessary further characterization techniques to identify mono- and few-layer flakes [7-9]. That is why techniques to thin down multilayered flakes and to deterministically transfer thin flakes have become highly desirable [10-15]. This is especially true for the case of very transparent 2D materials as the identification of these thin flakes can be challenging and very time consuming. For instance, single layer hexagonal boron nitride is very difficult to find due to its low optical absorption [16].

Here we employ an atomic force microscope tip attached to an XYZ manual micrometer stage to manipulate thick h-BN flakes. We find that the AFM tip can be used to tear the thick h-BN flakes, peeling off several layers. Single-layer h-BN flakes of several $\mu m^2$ areas can be produced with this method. The thinned down flakes are characterized using AFM and Raman spectroscopy.

**Materials**

Multilayer h-BN flakes are firstly deposited onto a $SiO_2$ (285 nm)/Si substrate by mechanical exfoliation of commercially available boron nitride powder (*Momentive, Polartherm* grade PT110) using blue Nitto tape (*Nitto Denko Co.,* SPV 224P). The AFM tips used in this work are *Nanosensors* ATEC-NC-10 which feature a 'visible tip'. These tips (conventionally used in tapping mode AFM) have a spring constant of ~40 N/m and tip radius of 20 nm. We find that blunt AFM tips (such as those discarded after an accidental tip-sample crash) are optimal for the manipulation and exfoliation process. When new, sharp AFM tips are used they typically cut through the boron nitride flakes (see Figure S1 in the Supporting Information).

**Experimental setup**

AFM based manipulation methods such as nanolithography, local oxidation/reduction, and scratching have been previously used to modify 2D materials [17-22]. Tip-induced exfoliation has also been demonstrated to thin down 2D materials but results in small areas [23]. Another drawback of previous tip-induced exfoliation is that it requires a complete AFM system. Here we demonstrate that an AFM tip attached to a manual XYZ micrometer stage is enough to manipulate and exfoliate atomically thin flakes, reducing considerably the cost and complexity of this technique.

Figure 1a shows a schematic diagram of the employed setup. We have slightly modified a setup employed to deterministically transfer 2D materials (described in detail in the Supporting Information of Ref. [13]). Briefly, the sample is fixed to a XY micrometer stage by double-sided tape and an AFM cantilever chip is attached (also with double-sided tape) to a glass slide which can be moved with an XYZ micromanipulator. A long working distance zoom lens system is used to accurately position the tip over the sample and to monitor the manipulation process. The tip-induced manipulation is carried out by employing the AFM tip as a micro-plowing tool, as depicted in Figure 1b.

**Results**

*Step-by-step microplowing process*

Figure 2 presents several frames captured during the manipulation and exfoliation of a thick (>50 nm) h-BN flake. The complete video can be found in the Supporting Information (or following this link) along with examples of additional flakes. In Figure 2.1 the h-BN is focused while the AFM tip is blurred because it is far above the surface. Then the flake is moved to the right until it is hidden by the cantilever. Note that arrow type cantilevers are used which have the tip located at the very end of the cantilever. The cantilever is slowly lowered until the tip is touching the surface of the sample. This can be detected because the reflectivity of the cantilever changes once the tip is in contact with the sample due to the deflection of the AFM cantilever (see Figure 2.3). After that, the sample is moved to the left (we move the sample stage because it is more stable than the tip stage). The AFM tip plows the surface, peeling off several layers (but not all) of the h-BN flake (Figure 2.4). The flake is, in this way, thinned down by exfoliating the topmost layers with the AFM tip. The reduction in thickness can be seen by the drastic change in color of the remaining flake. After the tip-induced exfoliation, the flake that has been peeled off is transferred to the surface nearby the original flake.

*Characterization of the plowed flakes*

Figure 3 presents the AFM characterization of the same h-BN flake thinned down in Figure 2. The optical image (Figure 3a) shows a clear change in the optical color of the plowed part with respect to the pristine part of the flake. An AFM topography image, acquired in the region marked by a square in Figure 3a, is shown in Figure 3b. A topographic line profile (measured along the dashed line in Figure 3b) is included below the AFM image. The thinner region has a thickness of ~0.7 nm, compatible with the thickness reported for monolayer h-BN in literature [16,24]. Moreover, from the analysis of the AFM image, the flake (originally >50 nm thick) has been thinned down to ~0.7 – 10 nm after the micro-plowing AFM process. The total area thinned down is >225 $\mu m^2$. We found that about 1/3 of the flakes can be thinned with large few-layer to monolayer areas. Therefore, this method can facilitate the production of monolayer h-BN flakes. This is especially important when powder h-BN is used as starting bulk material for exfoliation (sources of high quality h-BN single crystals are very scarce) as the density of thin flakes is typically very low.

Figure 4 shows another example of a thinned flake using the AFM micro-plowing technique which is further characterized using Raman spectroscopy. Figure 4a and 4b show the flake before (4a) and after (4b) micro-plowing. The 190 nm thick flake is reduced to a thickness of 5 nm after micro-plowing (see superimposed AFM line profiles in Figure 4b). Raman spectra of the $SiO_2$ and BN peaks at three locations after micro-plowing are shown in Figures 4c and 4d. The $SiO_2$ peak deceases in intensity as the BN flake becomes thicker. A complete Raman map of the reduced thickness flake and peeled flake is shown in Figure 4e. The color map shows the ratio of the intensity of the BN peak to that of the $SiO_2$ peak.

**Conclusions**

In summary, we have presented a method to manipulate and exfoliate h-BN flakes with an AFM tip. By using the tip as a micro-plowing tool, thick BN flakes can be peeled off leaving thinned down, large area flakes. We demonstrated that the process can be carried out by mounting the tip in a manual micrometer stage, without the need of a complete AFM system, therefore reducing the complexity and cost of the process. This technique can be applied to other 2D materials to manipulate thin flakes of different van der Waals crystals.

### Acknowledgements

The authors would like to acknowledge the FP7-Marie Curie Project PIEF-GA-2011-300802 ('STRENGTHNANO') and the Dutch organization for Fundamental Research on Matter (FOM).

**FIGURES**

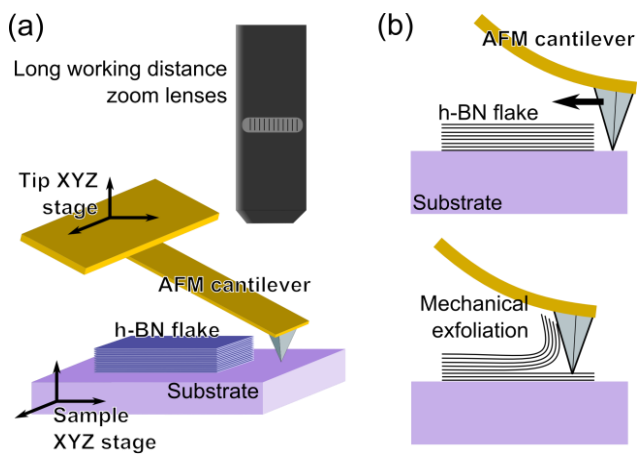

**Figure 1.** (a) Schematic of the experimental setup employed for the micro-plowing manipulation. Both the sample and an AFM cantilever are mounted in XYZ stages. A long working distance zoom lens is used to inspect the tip and sample during the process. (b) Diagram of the micro-plowing induced mechanical exfoliation. The AFM tip is lowered until it is in contact with the sample and then it is moved parallel to the sample, cleaving several layers of the boron nitride flake.

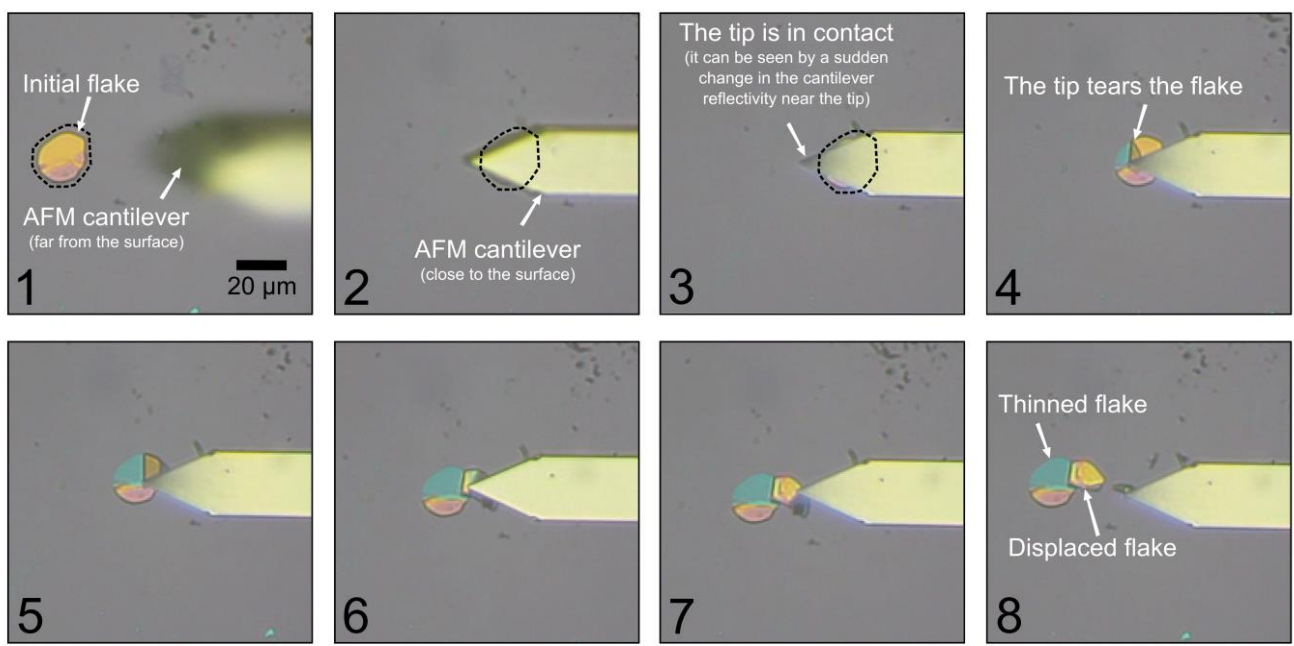

**Figure 2.** Frames captured during the micro-plowing based mechanical cleavage of a thick boron nitride flake. (1) The boron nitride flake is visible and the AFM tip is far from the surface. (2) The boron nitride flake is moved underneath the cantilever (which is lowered, almost touching the surface). (3) The cantilever is lowered even more until it is touching the surface. The change in reflectivity shows that it is in contact with the surface. (4) to (6) The sample is slowly displaced to the left. The AFM tip tears away the flake leaving behind a thinned boron nitride flake. The cleaved part of the flake is deposited nearby by grabbing it with the tip (7) to (8) (see Supplemental Information for complete video).

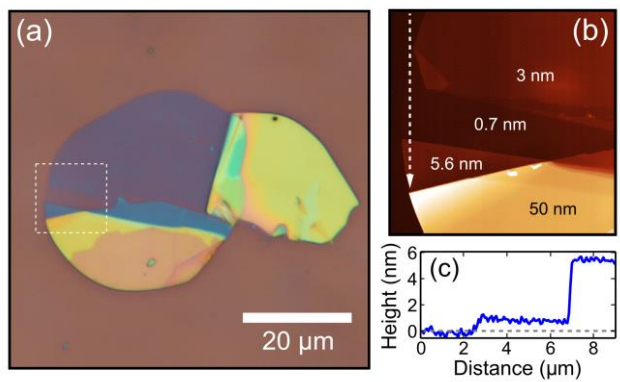

**Figure 3.** (a) Optical image of the micro-plowing exfoliated flake (from Figure 2) showing a big region which has been thinned down to few-layers. (b) AFM topography image of the region highlighted with a dashed white square. The original flake (about 50 nm thick) has been thinned down to 0.7 nm - 3 nm (~1 - 10 layers). (c) A topographic line profile is shown measured along the dashed line in (b).

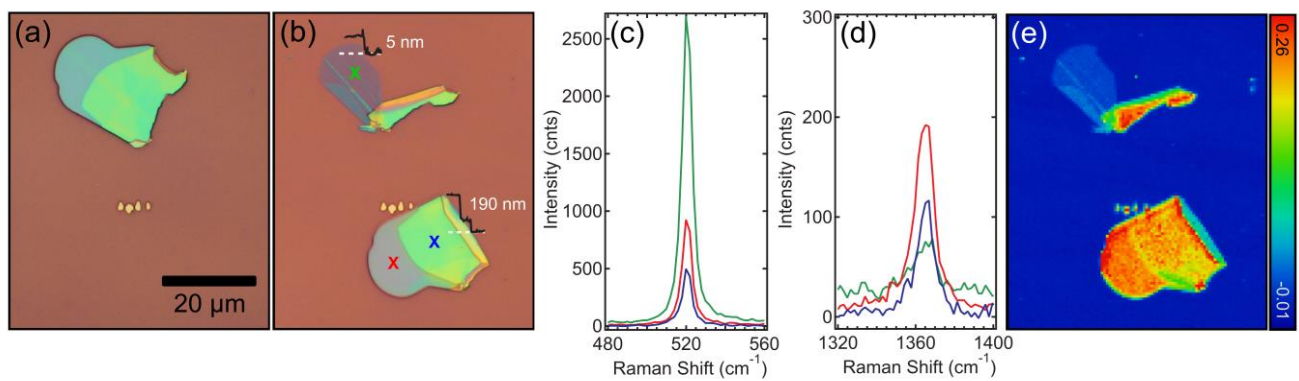

**Figure 4.** (a) Optical image of another thick h-BN flake before micro-plowing (b) Same flake in (a) after micro-plowing with an AFM tip. The 190 nm thick flake is exfoliated down to 5 nm (black

curves show superimposed AFM line profiles taken at the location of the dotted white lines). The green, red, and blue X's mark the locations of the Raman spectra in (c) and (d) having the same colors. (c) Raman spectra of the $SiO_2$ peak at 520 $cm^{-1}$ at the locations marked in (b). (d) Raman spectra of the BN peak at 1366 $cm^{-1}$ at the locations marked in (b). (e) Complete Raman map of the optical image in (b) showing the ratio of the intensities of the BN peak to that of the $SiO_2$ peak.

Supporting Information

# Mechanical manipulation and exfoliation of boron nitride flakes by micro-plowing with an AFM tip

*Joshua O. Island, Gary A. Steele, Herre S.J. can der Zant, Andres Castellanos-Gomez*[*]

**Supporting Information Contents**

1. **Effect of the tip sharpness on the manipulation and exfoliation**
    **Figure S1:** Cutting boron nitride flakes with sharp AFM tips

2. **Additional optical images of the micro-plowing process for two more flakes**
    **Figure S2:** Micro-plowing process for flake 1
    **Figure S3:** Micro-plowing process for flake 2

### 1. Effect of the tip sharpness on the manipulation and exfoliation

During this work we found that blunt AFM tips are optimal for the manipulation and exfoliation process. We employed AFM tips that were discarded after several hours of scanning which eventually yielded to tip crashes and severe reduction of the AFM resolution. When new, sharp AFM tips are used they typically cut through the boron nitride flakes (see Figure S1). In order to blunt the AFM tip one can put the tip in hard contact with the hard $SiO_2$/Si substrate and move the substrate in X and Y directions several times. After this process, the manipulation and exfoliation can be easily carried out without cutting the boron nitride flakes with the tip.

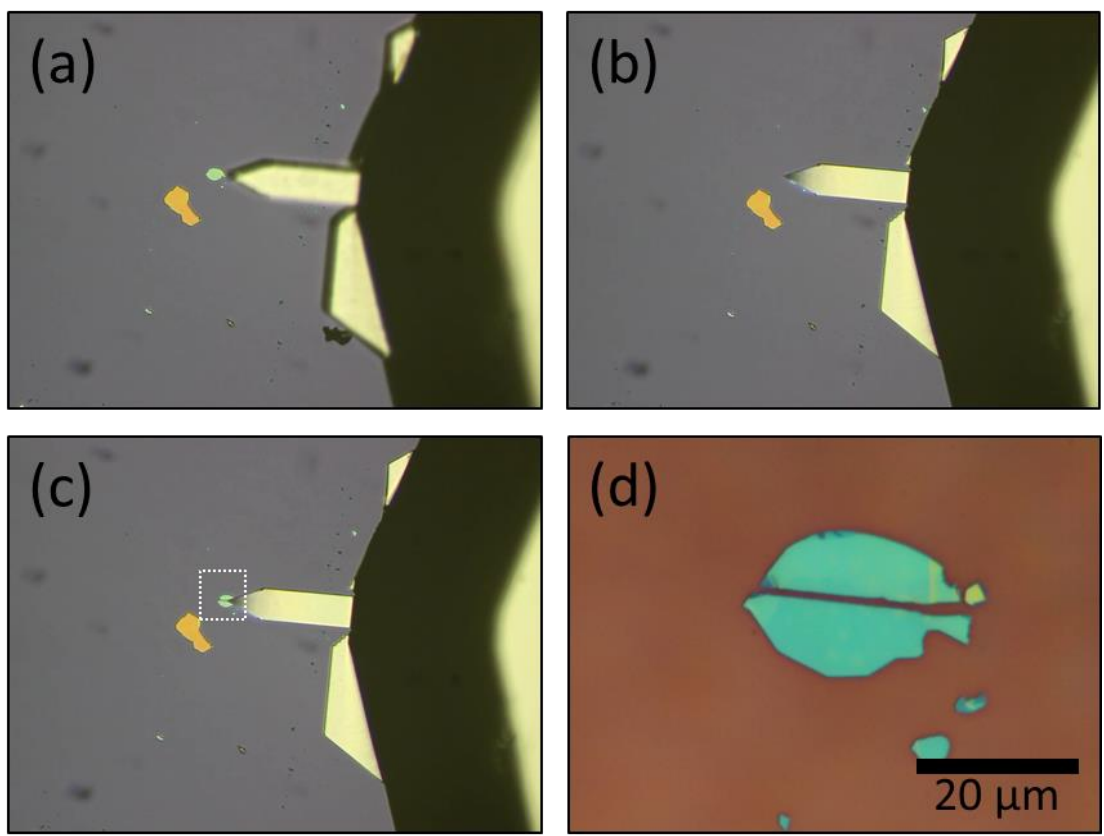

**Figure S1:** (a) A boron nitride flake is visible near a sharp new AFM tip. (b) The boron nitride flake is moved underneath the cantilever (which is lowered until touching the surface). (c) The sample is slowly displaced to the left using the stage. The sharp AFM tip cuts through the flake leaving a trench across the flake. (d) Optical image showing the cut made on the boron nitride flake by using a sharp AFM tip.

### 2. Additional optical images of the micro-plowing process for two more flakes

Here we show optical images of the micro-plowing process for two additional flakes. In all attempts, about 1/3 of the flakes can be significantly reduces in thickness from their starting thicknesses and in some cases even down to single layers (see main text). Figure S1 shows the complete micro-plowing process for flake 1, and Figure 2 shows the complete process for flake 2.

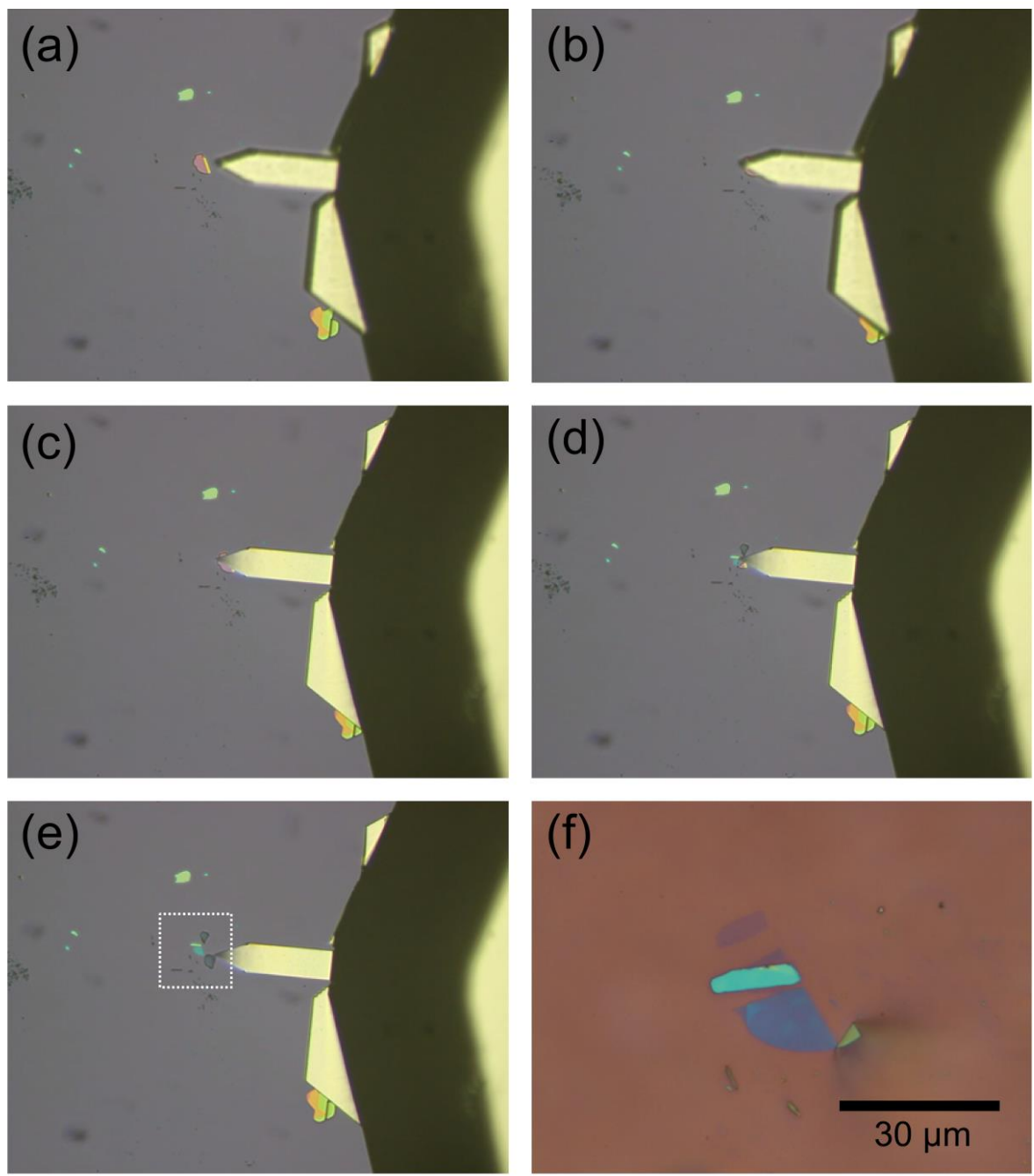

**Figure S2:** (a) The boron nitride flake is visible near the AFM tip. (b) The boron nitride flake is moved underneath the cantilever (which is lowered, almost touching the surface). (c) The cantilever is lowered even more until it is touching the surface. The change in reflectivity shows that it is in contact with the surface. (d) to (e) The sample is slowly displaced to the left using the stage. The AFM tip tears away the flake leaving behind a thinned boron nitride flake. (f) Optical image showing the thinned-down boron nitride flake.

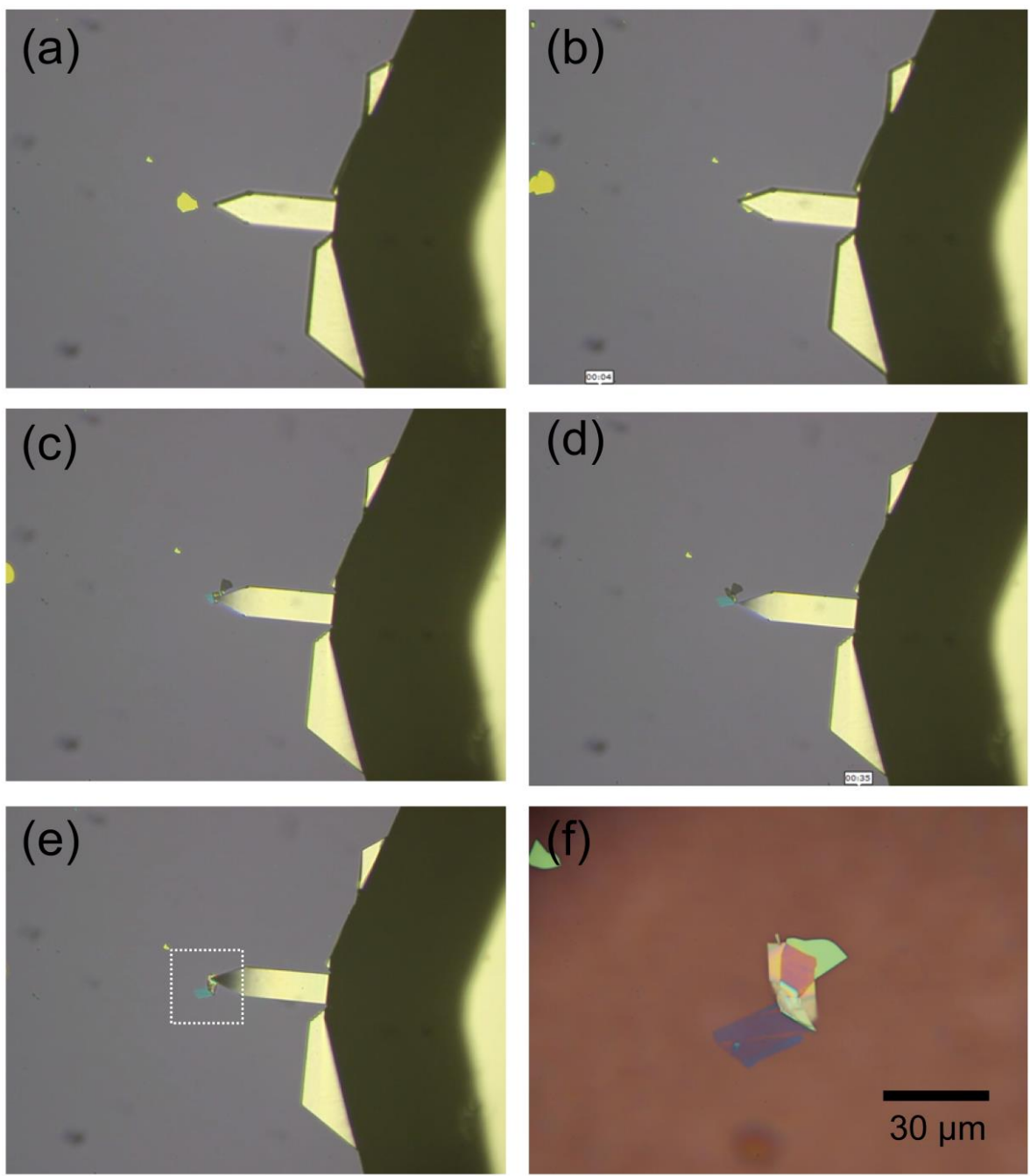

**Figure S3:** (a) The boron nitride flake is visible near the AFM tip. (b) The boron nitride flake is moved underneath the cantilever (which is lowered, almost touching the surface). (c) The cantilever is lowered even more until it is touching the surface. The change in reflectivity shows that it is in contact with the surface. (d) to (e) The sample is slowly displaced to the left using the stage. The AFM tip tears away the flake leaving behind a thinned boron nitride flake. (f) Optical image showing the thinned-down boron nitride flake.